\documentclass[twocolumn,superscriptaddress,amssymb,aps,pra]{revtex4-2}
\usepackage{amsmath,amsfonts,amssymb,amsthm,graphics,graphicx,epsfig,bbm}
\usepackage[colorlinks=true,citecolor=blue,linkcolor=blue,urlcolor=blue]{hyperref}
\usepackage[usenames]{color}

\usepackage{dsfont}
\usepackage{txfonts}
\usepackage{amsfonts}
\usepackage{mathrsfs}
\usepackage{physics}
\usepackage{color}
\usepackage{epstopdf}
\usepackage{bbold}
\usepackage{mathtools}
\usepackage{multirow}
\usepackage{subfigure}

\begin{document}

\title{Protecting quantum gates from arbitrary single- and two-qubit errors}

\author{Chunfeng Wu}
\email{chunfeng\_wu@sutd.edu.sg}
\affiliation{Science, Mathematics and Technology, Singapore University of Technology and Design, 8 Somapah Road, Singapore 487372, Singapore}

\author{Gangcheng Wang}
\affiliation{School of Physics and Center for Quantum Sciences, Northeast Normal University, Changchun 130024, China}

\author{Xun-Li Feng}
\affiliation{Science, Mathematics and Technology, Singapore University of Technology and Design, 8 Somapah Road, Singapore 487372, Singapore}

\begin{abstract}
We explore the protection of quantum gates from arbitrary single- and two-qubit noises with properly designed dynamical decoupling pulses. The proposed dynamical decoupling method is a concatenation of a sequence of pulses formed by $\sigma_x$, $\sigma_x\sigma_x$ with another sequence constructed by $\sigma_z$, $\sigma_z\sigma_z$. The concatenation of the two sequences results in desired pulses to fight agianst any single- and two-qubit errors. The success of our method relies on the ability to adjust system parameters or interaction terms, which can be achieved in different physical systems, including trapped ions and superconducting qubits. We finally explore the performance of our method numerically with the above-mentioned errors that are changing at any moment and show the preferred protection offered by the method. Therefore, our method is a timely step forward in preserving quantum gates at the level of physical qubits.

\end{abstract}

\date{\today}

\maketitle

Fault-tolerant quantum computation requires to scale up to enough qubits and implement quantum gates with sufficiently small errors in physical systems \cite{shor96, aharnov97, Kitaev97, Zurek98, Ruskai00, Hollenberg11}. These significant challenges are expected to be overcome step by step with growing noisy intermediate small-scale quantum (NISQ) devices \cite{nisq1,nisq2}. To realize scalable quantum computation in NISQ devices, the system interactions should be adjustable in order to execute quantum gates on target qubits. The desired controllability in system parameters may induce additional difficulties besides decoherence in manipulating the system evolution for achieving specific quantum gates, because of unavoidable fluctuations in the systerm parameters \cite{twoq1, twoq2}.

Dynamical decoupling (DD) plays an effective role in mitigating errors or decoherence during system evolution \cite{dd1, dd2, dd3, dd4, dd5}. The technique aims to approximately kick out errors or decoherence via suitable pulses applied in sequence under certain conditions. Compared with quantum error-correction codes, DD is also an active method to battle with errors or decoherence but with a moderate resouce of DD pulses. This merit makes DD practically useful in NISQ devices. In commonly explored physical systems, most of the errors during system evolution are due to single- and two-qubit noises when qubit-qubit couplings are demanded. In the literature, fighting against aribitrary single-qubit noises by DD approach has been extensively investigated both theoretically and experimentally \cite{dd1, dd2, dd3, dd4, dd5}. Meanwhile, it has been directed towards mitigating arbitrary two-qubit noises with DD technique in physical systems as well \cite{twoq0, twoq02, twoq1, twoq2, twoq3}. In Refs. \cite{twoq0,twoq02}, it was shown that system operators can generally be protected by complicated levels of nested DD pulses, but without discussing the details of eliminating arbitrary two-qubit errors or decoherence in physical systems. In 2016, the Viola group explored the supression of multi-qubit dephasing by comparatively short DD sequences \cite{twoq1}. Generally, the DD techniques cannot be readily implemented in actual experiments since the DD pulses may remove desired system interactions when fighting against errors or decoherence. It is thus of practical importance to make the DD techniques workable in actual experments to battle with both single- and two-qubit errors. Recently, special DD pulses invovling two-qubit interactions were designed to minimize the negative effect of specific two-qubit errors by adjusting system parameters in a system of superconducting transmon qubits in Ref. \cite{twoq2}. One year later, the Lidar group investigated the mitigation of ZZ coupling in superconducting qubits \cite{twoq3, twoq4} and the Kim group studied the supression of typical crosstalk in implementing the M\o lmer-S\o rensen (MS) gate with trapped ions \cite{twoq5}.

In this work, we explore the supression of arbitrary single- and two-qubit errors or decoherence through DD approach with experimental achievable Hamiltonians. Generally, arbitrary single- and two-qubit errors are described by the error operators in the set of $\mathcal{E}=\{\sigma_{x,y,z}^{i,j},\sigma_x^i\sigma_x^j,\sigma_x^i\sigma_y^j,\sigma_y^i\sigma_x^j,\sigma_x^i\sigma_z^j,\sigma_z^i\sigma_x^j,\sigma_y^i\sigma_y^j,\sigma_y^i\sigma_z^j,\sigma_z^i\sigma_y^j,\sigma_z^i\sigma_z^j\}$. We first study one DD sequence constructed by $\sigma_x^{i,j}$ and illustrate that the errors in the subset $\mathcal{E}_1=\{\sigma_{y,z}^{i}$, $\sigma_{y,z}^{j}$, $\sigma_x^i\sigma_y^j$, $\sigma_y^i\sigma_x^j$, $\sigma_x^i\sigma_z^j$, $\sigma_z^i\sigma_x^j$, $\sigma_y^i\sigma_y^j$, $\sigma_y^i\sigma_z^j$, $\sigma_z^i\sigma_y^j$, $\sigma_z^i\sigma_z^j\}$ can be eliminated approximately. Next we investigate another DD sequence built by $\sigma_z^{i,j}$ to supress the errors in the subset of $\mathcal{E}_2=\{\sigma_x^{i,j},\sigma_x^i\sigma_x^j\}$. The concatenation of the two DD sequences can thus fight against any errors in the full set as $\mathcal{E}=\mathcal{E}_1+\mathcal{E}_2$, and this result is consistent with the
nested Uhrig dynamical decoupling (NUDD) explored in Ref. \cite{twoq0,twoq02}. During the process of applying DD pulses, system parameters should be controllable such that required Hamiltonian terms will not be affected by the DD pulses. The requirement is experimentally achievable in commonly studied physical systems and as a result, a universal set of protected quantum gates are realizable with current experimental techniques based on our method. We next explore the performance of our method by modelling all the above-mentioned errors to be time-dependent in numerical calculations. The perfomance is dependent on the number of DD pulses and the more the number of DD pulses, the better the performance is. But in practical experiments, the number of DD pusles is finite. We then show with a moderate number of repeating the concatenated DD sequences, the explored quantum gates can be largely saved from the errors. It is reasonable to expect that the performance can be further improved given more DD pusles with the development of experimental techniques.

We study the first DD sequence constructed by $\sigma_x^{i,j}$ to fight against the errors included in the subset $\mathcal{E}_1$. The desired sequence is of the form, $\big(\sigma_x^j[\cdot]\sigma_x^j\big)\big(\sigma_x^i[\cdot]\sigma_x^i\big)\big(\sigma_x^i\sigma_x^j[\cdot]\sigma_x^i\sigma_x^j\big)[\cdot]$, where $[\cdot]$ represents evolution operator over a period of $\tau$. We analyze the system evolution with DD pusles step by step in the following.
\begin{itemize}
\item The evolution $[\cdot]$ carries the errors as $\{\sigma_{y,z}^{i}$, $\sigma_{y,z}^{j}$, $\sigma_x^i\sigma_y^j$, $\sigma_y^i\sigma_x^j$, $\sigma_x^i\sigma_z^j$, $\sigma_z^i\sigma_x^j$, $\sigma_y^i\sigma_y^j$, $\sigma_y^i\sigma_z^j$, $\sigma_z^i\sigma_y^j$, $\sigma_z^i\sigma_z^j\}$.
\item $\big(\sigma_x^i\sigma_x^j[\cdot]\sigma_x^i\sigma_x^j\big)$ converts the errors to $\{-\sigma_{y,z}^{i}$, $-\sigma_{y,z}^{j}$, $-\sigma_x^i\sigma_y^j$, $-\sigma_y^i\sigma_x^j$, $-\sigma_x^i\sigma_z^j$, $-\sigma_z^i\sigma_x^j$, $\sigma_y^i\sigma_y^j$, $\sigma_y^i\sigma_z^j$, $\sigma_z^i\sigma_y^j$, $\sigma_z^i\sigma_z^j\}$.
\item $\big(\sigma_x^i[\cdot]\sigma_x^i\big)$ converts the errors to $\{-\sigma_{y,z}^{i}$, $\sigma_{y,z}^{j}$, $\sigma_x^i\sigma_y^j$, $-\sigma_y^i\sigma_x^j$, $\sigma_x^i\sigma_z^j$, $-\sigma_z^i\sigma_x^j$, $-\sigma_y^i\sigma_y^j$, $-\sigma_y^i\sigma_z^j$, $-\sigma_z^i\sigma_y^j$, $-\sigma_z^i\sigma_z^j\}$.
\item $\big(\sigma_x^j[\cdot]\sigma_x^j\big)$ converts the errors to $\{\sigma_{y,z}^{i}$, $-\sigma_{y,z}^{j}$, $-\sigma_x^i\sigma_y^j$, $\sigma_y^i\sigma_x^j$, $-\sigma_x^i\sigma_z^j$, $\sigma_z^i\sigma_x^j$, $-\sigma_y^i\sigma_y^j$, $-\sigma_y^i\sigma_z^j$, $-\sigma_z^i\sigma_y^j$, $-\sigma_z^i\sigma_z^j\}$.
\end{itemize}
As a whole, $\big(\sigma_x^j[\cdot]\sigma_x^j\big)\big(\sigma_x^i[\cdot]\sigma_x^i\big)\big(\sigma_x^i\sigma_x^j[\cdot]\sigma_x^i\sigma_x^j\big)[\cdot]$ removes the errors in the subset $\mathcal{E}_1$ approximately. To eliminate the errors in the subset $\mathcal{E}_2$, we utilize the second DD sequence in the form of $\big(\sigma_z^j[\cdot]\sigma_z^j\big)\big(\sigma_z^i[\cdot]\sigma_z^i\big)\big(\sigma_z^i\sigma_z^j[\cdot]\sigma_z^i\sigma_z^j\big)[\cdot]$. Only with the errors in the subset $\mathcal{E}_2$, we find
\begin{itemize}
\item The evolution $[\cdot]$ carries the errors as $\{\sigma_x^{i,j}$, $\sigma_x^i\sigma_x^j\}$.
\item $\big(\sigma_z^i\sigma_z^j[\cdot]\sigma_z^i\sigma_z^j\big)$ converts the errors to $\{-\sigma_x^{i,j}$, $\sigma_x^i\sigma_x^j\}$.
\item $\big(\sigma_z^i[\cdot]\sigma_z^i\big)$ converts the errors to $\{-\sigma_x^{i}$, $\sigma_x^{j}$, $-\sigma_x^i\sigma_x^j\}$.
\item $\big(\sigma_z^j[\cdot]\sigma_z^j\big)$ converts the errors to $\{\sigma_x^{i}$, $-\sigma_x^{j}$, $-\sigma_x^i\sigma_x^j\}$.
\end{itemize}
Therefore, the errors in the subset $\mathcal{E}_2$ can be effectively mitigated by the second DD sequence roughly. Finally, the concatenation of the two DD sequences gives us $\big(\sigma_z^j[\cdot\cdot]\sigma_z^j\big)\big(\sigma_z^i[\cdot\cdot]\sigma_z^i\big)\big(\sigma_z^i\sigma_z^j[\cdot\cdot]\sigma_z^i\sigma_z^j\big)[\cdot\cdot]$, where $[\cdot\cdot]=\big(\sigma_x^j[\cdot]\sigma_x^j\big)\big(\sigma_x^i[\cdot]\sigma_x^i\big)\big(\sigma_x^i\sigma_x^j[\cdot]\sigma_x^i\sigma_x^j\big)[\cdot]$. The nested DD sequence can approximately get rid of all the errors in the set $\mathcal{E}=\mathcal{E}_1+\mathcal{E}_2$. Moreover, we can reduce the number of DD pulses by simplifying the nested DD sequence as $\sigma_z^j[\cdot\cdot]\sigma_z^j\sigma_z^i[\cdot\cdot]\sigma_z^j[\cdot\cdot]\sigma_z^i\sigma_z^j[\cdot\cdot]$, where $[\cdot\cdot]=\sigma_x^j[\cdot]\sigma_x^j\sigma_x^i[\cdot]\sigma_x^j[\cdot]\sigma_x^i\sigma_x^j[\cdot]$. It is clearly shown that to complete one full DD cycle, there are 16 steps of system evolution and so, the total evolution time is $T=16\tau$.

Our scheme is possibly implementable in physical systems, with tunable system parameters such that desired gate Hamiltonians will not be canceled with DD pulses applied. One possible gate Hamiltonian is of the below form,
\begin{eqnarray} \label{h1}
H_1=\delta \sigma_z^j  + \Omega (e^{-i\phi}\sigma_+^j + e^{i\phi}\sigma_-^j) + J(\sigma_+^j \sigma_-^k+\sigma_-^j \sigma_+^k),
\end{eqnarray}
where  $\delta$ is the detuning of the qubit frequency from the frequency of external driving, $\Omega$ is the Rabi frequency of external driving, $\phi$ is the phase of external driving, $\sigma_{\pm}^{j,k} =\frac{1}{2}(\sigma_x^{j,k}\pm i\sigma_y^{j,k})$, and $J$ is the qubit-qubit coupling strength. The desired interaction can be implemented in various physical systems with rotating-wave approximation applied, including cold atoms~\cite{cold}, trapped ions~\cite{ion1, ion2} and superconducting transmon qubits \cite{trans1,trans2,trans3}, etc. The parameters $\delta$, $\Omega$, $\phi$ and $J$ are adjustable experimentally.

With gate Hamiltonian (\ref{h1}), single-qubit rotations about $x$ and $y$ axes can be achieved by choosing $\delta=0$, $J=0$ and tuning $\phi=0$ or $\pi$ for $x$ axis and $\phi=\frac{\pi}{2}$ or $\frac{3\pi}{2}$ for $y$ axis. While single-qubit rotations about $z$ axis is realizable by making $\Omega=0$ and $J=0$. For two-qubit entangling gates, one can select $\delta=0$ and $\Omega=0$ to keep the coupling term only.

In what follows, we consider a physical system of transmon qubits to demonstrate the implementation of quantum gates protected by our scheme against arbitrary single- and two-qubit errors. We study one unmodulated qubit (qubit 1) coupled to another frequency-modulated qubit (qubit 2) with external drives. The Hamiltonian in the interaction picture can be written as~\cite{trans1,trans2,trans3},
\begin{eqnarray}
H_{\rm trans}&=&\delta \sigma_z^{1,2} +\Omega (e^{-i\phi}\sigma_+^{1,2} + e^{i\phi}\sigma_-^{1,2}) \nonumber\\
&+& gJ_1(\beta)(\sigma_+^1 \sigma_-^2e^{-i\varphi}+\sigma_-^1 \sigma_+^2e^{i\varphi}),
\end{eqnarray}
where $g$ is the coupling strength, $J_1(\beta)$ is the Bessel function of the first kind when $n=1$, $\beta$ is the ratio of amplitude and frequency of the periodic frequency modulation, and $\varphi$ is the phase of the periodic frequency modulation. The $H_{\rm trans}$ is of the same form as $H_1$, in which $J=gJ_1(\beta)$.

In the system, the performance of the quantum gates proteced by our DD scheme can be demonstrated by the following numerical results. In the presence of errors or decoherence, we assume the total Hamiltonian is written as $H_T=H_S+H_e$, where $H_S$ and $H_e$ are system Hamiltonian and the stochastic error term. In our calculations, $H_S$ is $H_{\rm trans}$ and $H_e$ is given by
\begin{eqnarray}
H_{e}&=&\delta_{1,2}^x\sigma_x^{1,2}+\delta_{1,2}^y\sigma_y^{1,2}+\delta_{1,2}^z\sigma_z^{1,2}+\delta_{12}^{xx}\sigma_x^1\sigma_x^2+\delta_{12}^{yy}\sigma_y^1\sigma_y^2\nonumber\\
&+&\delta_{12}^{zz}\sigma_z^1\sigma_z^2+\delta_{12}^{xy}\sigma_x^1\sigma_y^2+\delta_{12}^{yx}\sigma_y^1\sigma_x^2+\delta_{12}^{xz}\sigma_x^1\sigma_z^2+\delta_{12}^{zx}\sigma_z^1\sigma_x^2\nonumber\\
&+&\delta_{12}^{yz}\sigma_y^1\sigma_z^2+\delta_{12}^{zy}\sigma_z^1\sigma_y^2,
\end{eqnarray}
where $\delta_{1,2}^{x,y,z}$ and $\delta_{12}^{uv} (u,v=x,y,z)$ describe the time-dependent strength of stochastic errors in different directions and couplings. To model the time-dependance, we create random numbers with respect to time (800 sets of values over a period of $T$) from a uniform distribution $[2\pi\times 1, 2\pi\times 10]$ MHz when solving the differential equation with ode45 solver.

For single-qubit gates, it is easy to apply the protection by DD approach as we only need to manipulate one qubit. According to the literature~\cite{twoq3, twoq4}, XY4 or Periodic DD (PDD) sequence is able to remove all errors described by $\sigma_{x,y,z}$ on the qubit with DD pusles applied. Therefore, it is not necessary to use the DD scheme proposed here since it is more complicated. We focus on fighting against errors or decoherence when exploring the implementation of two-qubit gates. To realize an entangling gate $U_3=e^{-i\gamma(\sigma_+^1 \sigma_-^2+\sigma_-^1 \sigma_+^2)}$ (where $\gamma=JT=\pi/4$), we select system parameter as $J=2\pi\times 10$ MHz. For a two-qubit gate, XY4 or PDD is not able to remove all errors any more and our DD scheme takes turn. Some of the DD pulses change desired interaction from $\sigma_+^1 \sigma_-^2+\sigma_-^1 \sigma_+^2$ to $\sigma_+^1 \sigma_+^2+\sigma_-^1 \sigma_-^2$ and some of the DD pusles change $J$ to $-J$. The changes certainly ruin the implementation of the entangling gates. Therefore, we need to tune the coupling term correspondingly to avoid the unwanted changes. In the former case, we control the frequency modulation to get $\sigma_+^1 \sigma_+^2+\sigma_-^1 \sigma_-^2$, which will be changed to desired interaction with certain DD pusles applied. In the latter case, we adjust $\varphi$ to obtain a negative value of $J$, which will then be converted back to $J$ by some DD pulses. We list out required coupling in Table \ref{t1} (a) for all the 16 steps of system evolution. In Table \ref{t1} (a), we use the symbol ``*" to indicate at that step, the coupling is set to be $\sigma_+^1 \sigma_+^2+\sigma_-^1 \sigma_-^2$ and without the symbol, the coupling is of the form $\sigma_+^1 \sigma_-^2+\sigma_-^1 \sigma_+^2$. With the specified couplings, our DD scheme is workable in the physical system as demonstrated by the following numerical calculations. We randomly choose 50 initial states and calculate average fidelity of the gate when $\gamma=\pi/4$ in the absence or presence of errors (without DD and with DD), respectively. Specially, we simluate the DD pulses by unitary evolutions over a short time period and take errors occurred to DD pulses into account in our calculations with $\sigma_x\propto e^{-i(\frac{\pi}{2}+\zeta)\sigma_x}$ and $\sigma_z\propto e^{-i(\frac{\pi}{2}+\zeta)\sigma_z}$, where $\zeta$ is selected from Gaussian distribution (1) with a mean of $\frac{\pi}{500}$ and a standard deviation of $\frac{\pi}{500}$ and from Gaussian distribution (2) with a mean of $\frac{\pi}{200}$ and a standard deviation of $\frac{\pi}{200}$ to describe different stochastic errors happened to the DD pulses. The numerical results are summarized in Table \ref{t2}. Here, we consider the the case of getting $\sigma_+^1 \sigma_-^2+\sigma_-^1 \sigma_+^2$ and $\sigma_+^1 \sigma_+^2+\sigma_-^1 \sigma_-^2$ ideally by rotating-wave approximation. If include the approximation errors, the fideilities will be a bit lower. It is clearly demonstrated in Table \ref{t2}, the two-qubit gate is ruined by the errors without DD pulses applied, and it can be largely saved according to our DD scheme. Moreover, the success of the DD scheme requires very small errors happened in DD pulses, see numerical results in the case with superscript (1). Given the DD pulses with increased value of $\zeta$ as shown in the case with superscript (2), the DD scheme still can offer protection against the errors, though it does not perform as well as it does in the case with superscript (1). Here, we only complete one full DD cycle with 16 steps. Further improvement of the average fidelity can be achieved by increasing the number of repeating the DD cycle.

\begin{table}[h]
\begin{subtable}
\centering
\begin{tabular}{|c|c|}
\hline\hline
Steps& $U_3$   \\
\hline
1st $[\cdot]$  & $J$\\ \hline
2nd $[\cdot]$ & $J$\\ \hline
3rd $[\cdot]$ & $J*$\\ \hline
4th $[\cdot]$ & $J*$\\ \hline
5th $[\cdot]$ & $J$\\ \hline
6th $[\cdot]$ & $J$\\ \hline
7th $[\cdot]$ & $J*$\\ \hline
8th $[\cdot]$ & $J*$\\ \hline
9th $[\cdot]$ & $-J$\\ \hline
10th $[\cdot]$ & $-J$\\ \hline
11th $[\cdot]$ & $-J*$\\ \hline
12th $[\cdot]$ & $-J*$\\ \hline
13th $[\cdot]$ & $-J$\\ \hline
14th $[\cdot]$ & $-J$\\ \hline
15th $[\cdot]$ & $-J*$\\ \hline
16th $[\cdot]$ & $-J*$\\ \hline
\hline
\multicolumn{2}{c}{(a)}\\
\end{tabular}
\end{subtable}
\begin{subtable}
\centering
\begin{tabular}{|c|c|}
\hline\hline
Steps& $U_e^1$   \\
\hline
1st $[\cdot]$  & $J'$\\ \hline
2nd $[\cdot]$ & $J'$\\ \hline
3rd $[\cdot]$ & $-J'$\\ \hline
4th $[\cdot]$ & $-J'$\\ \hline
5th $[\cdot]$ & $J'$\\ \hline
6th $[\cdot]$ & $J'$\\ \hline
7th $[\cdot]$ & $-J'$\\ \hline
8th $[\cdot]$ & $-J'$\\ \hline
9th $[\cdot]$ & $J'$\\ \hline
10th $[\cdot]$ & $J'$\\ \hline
11th $[\cdot]$ & $-J'$\\ \hline
12th $[\cdot]$ & $-J'$\\ \hline
13th $[\cdot]$ & $J'$\\ \hline
14th $[\cdot]$ & $J'$\\ \hline
15th $[\cdot]$ & $-J'$\\ \hline
16th $[\cdot]$ & $-J'$\\ \hline
\hline
\multicolumn{2}{c}{(b)}\\
\end{tabular}
\end{subtable}
\begin{subtable}
\centering
\begin{tabular}{|c|c|}

\hline\hline
Steps& $U_e^2$   \\
\hline
1st $[\cdot]$  & $J'$\\ \hline
2nd $[\cdot]$ & $J'$\\ \hline
3rd $[\cdot]$ & $J'$\\ \hline
4th $[\cdot]$ & $J'$\\ \hline
5th $[\cdot]$ & $J'$\\ \hline
6th $[\cdot]$ & $J'$\\ \hline
7th $[\cdot]$ & $J'$\\ \hline
8th $[\cdot]$ & $J'$\\ \hline
9th $[\cdot]$ & $-J'$\\ \hline
10th $[\cdot]$ & $-J'$\\ \hline
11th $[\cdot]$ & $-J'$\\ \hline
12th $[\cdot]$ & $-J'$\\ \hline
13th $[\cdot]$ & $-J'$\\ \hline
14th $[\cdot]$ & $-J'$\\ \hline
15th $[\cdot]$ & $-J'$\\ \hline
16th $[\cdot]$ & $-J'$\\ \hline
\hline
\multicolumn{2}{c}{(c)}\\
\end{tabular}
\end{subtable}
\begin{subtable}
\centering
\begin{tabular}{|c|c|}

\hline\hline
Steps& $U_e^3$   \\
\hline
1st $[\cdot]$  & $J'$\\ \hline
2nd $[\cdot]$ & $-J'$\\ \hline
3rd $[\cdot]$ & $-J'$\\ \hline
4th $[\cdot]$ & $J'$\\ \hline
5th $[\cdot]$ & $-J'$\\ \hline
6th $[\cdot]$ & $J'$\\ \hline
7th $[\cdot]$ & $J'$\\ \hline
8th $[\cdot]$ & $-J'$\\ \hline
9th $[\cdot]$ & $J'$\\ \hline
10th $[\cdot]$ & $-J'$\\ \hline
11th $[\cdot]$ & $-J'$\\ \hline
12th $[\cdot]$ & $J'$\\ \hline
13th $[\cdot]$ & $-J'$\\ \hline
14th $[\cdot]$ & $J'$\\ \hline
15th $[\cdot]$ & $J'$\\ \hline
16th $[\cdot]$ & $-J'$\\ \hline
\hline
\multicolumn{2}{c}{(d)}\\
\end{tabular}
\end{subtable}
\caption{Required system parameters for different quantum gates at every step of system evolution to make the DD scheme compatible, where (a) is for gate Hamiltonian (\ref{h1}) for two-qubit gates with our DD, (b) is for gate Hamiltonian (\ref{h2}) with ZZ coupling, (c) is for the gate Hamiltonian (\ref{h2}) with XX coupling, and (d) is for gate Hamiltonian (\ref{h2}) with ZX coupling} \label{t1}
\end{table}

\begin{table}[ht]
\begin{tabular}{|c|c|c|c|}
\hline\hline
\multirow{3}{*}{Gate}  &  \multicolumn{3}{c|}{Average Fidelity}\\
\cline{2-4}
& w/o DD (w/ iDD) & w/o DD (w/ DD$^{(1)})$ & w/o DD (w/ DD$^{(2)})$\\
\hline
$U_3|_{\gamma=\pi/4}$  &$ 0.2496\; (0.9967)$ & $ 0.2290\; (0.9949)$  & $ 0.2282\; (0.9807)$\\ \hline
$U_e^1|_{\gamma=\pi/4}$ &$ 0.1808\; (0.9923)$ & $ 0.1602\; (0.9886)$  & $ 0.1519 \; (0.9776)$\\ \hline
\hline
\end{tabular}
  \caption{The fidelities of two-qubit quantum gates in the absence (presence) of errors and DD pulses with Hamiltonian $H_{\rm trans}$, where iDD is short for ideal DD, and  superscript (1) or (2) indicates the case with $\zeta$ selected from Gaussian distribution (1) or (2), correspondingly.} \label{t2}
\end{table}

Other possible gate Hamiltonians with qubit-qubit coupling only are given as
\begin{eqnarray} \label{h2}
H_2=J'\sigma_z^j \sigma_z^k \; \text{or}\; J'\sigma_x^j \sigma_x^k \; \text{or}\; J'\sigma_z^j \sigma_x^k,
\end{eqnarray}
where $J'$ is the coupling strength. Here, we neglect the terms for realizing single-qubit gates as they can be preserved by XY4 or PDD. The type of interaction can be found in superconducting charge qubits~\cite{cha1,cha2},  transmons~\cite{trans4, trans5} and fluxoniums~\cite{fluxo1, fluxo2}, etc. From gate Hamiltonian (\ref{h2}), two-qubit entangling gates $U_e^i$ ($i=1,2,3$ and $e$ stands for entangling) are achievable dependent on the interaction $J'\sigma_z^j \sigma_z^k$ or $J'\sigma_x^j \sigma_x^k$ or $J'\sigma_z^j \sigma_x^k$. Based on the results in the literature, the system parameters may be fully or partially controllable in different systems. To make our DD scheme compatible in experiments, we require the tunablity of the system parameters from positive to negative. Similarly, we list out desired system parameters to keep gate Hamiltonian (\ref{h2}) for two-qubit gates with DD pulses applied in Table \ref{t1} (b), (c) and (d), where (b) is for interaction $J'\sigma_z^j \sigma_z^k$, (c) is for interaction $J'\sigma_x^j \sigma_x^k$ and (d) is for interaction $J'\sigma_z^j \sigma_x^k$. In the table, $U_e^1=e^{-i\xi \sigma_z^1 \sigma_z^2}$, $U_e^2=e^{-i\xi \sigma_x^1 \sigma_x^2}$ and $U_e^3=e^{-i\xi \sigma_z^1 \sigma_x^2}$, where $\xi$ is dependent of the value of $J'$ and evolution time. The coupling term may take one of the forms or include two of them in some physical systems, because of crosstalk. As our DD scheme can diligently eliminate aribrary two-qubit errors, unwanted crosstalk term can also be kicked out by DD pulses without changing the sign of the coupling strength of the crosstalk term. It may happen that in some physical system, two-qubit gates cannot be executable in a protective way according to our scheme with current experimental techniques when there is a lack of desired tunability in the coupling strength. We then aniticipate further development from the perspective of experiments.

We next consider ZZ coupling as an example to explore the protection offered by our DD scheme through numerical calculations. The action of DD pulses sometimes changes the sign of the ZZ coupling strength. To remain the interaction for realizing a two-qubit gate, we need to adjust the sign as demonstrated in Table \ref{t1} (b). In the example, we investigate the average fidelity of gate $U_e^1$ with $\xi=J'T=\pi/4$ and $J'=2\pi\times 10$ MHz by randomly choosing 50 initial states, without errors and with errors (without DD and with DD). The numerical results are obtained by completing one whole DD cycle of 16 steps, and are outlined in Table \ref{t2} too. We find the two-qubit gate is in ruins with the errors in the absence of DD pulses and it can be preserved when DD pusles are present. Similarly, the performance of the DD scheme relies on desirably small errors happened to the DD pulses, described by the value of $\zeta$ and the preservation offered is expected to be enhanced if we repeat the whole DD cycle multiple times.

To summarize, we present one DD scheme to battle with arbitrary single- and two-qubit errors. The DD scheme is based on the concatenation of the two DD sequences. To make the DD scheme compatible with realizing quantum gates in physical systems, we require tunable system parameters such that desired gate Hamiltonians will not be eliminated by the DD pulses. We investigate the implementation of the DD scheme based on gate Hamiltonian (\ref{h1}), which can be achieved experimentally in cold atoms, trapped ions and superconducting qubits with controllable system parameters. Specifically, we demonstrate that two-qubit quantum gates are excellently preserved against arbitrary single- and two-qubit errors by considering a physical system of transmon qubits. Moreover, we also show that the DD scheme is workable with gate Hamiltonian (\ref{h2}), that is experimentally attainable in superconducting charge qubits, flux qubits and fluxoniums. With the current experimental techniques, however, it may not be possible to obtain excellently protected quantum gates since some of the parameters in (\ref{h2}) may not be adjustable as desired by the DD scheme. We expect to see further development in actual experiments to realize controllable parameters in near future. Our DD scheme provides an efficient and practical way to mitigate arbitrary single- and two-qubit errors that are the main source of noises in executing quantum gates, and so it is a timely contribution to improving NISQ devices.

\vspace{5pt}
This research is supported by the National Research Foundation, Singapore and A*STAR under its Quantum Engineering Programme (NRF2021-QEP2-02-P03). G.W. is supported by the Fundamental Research Funds for the Central Universities (Grant No. 2412020FZ026).

\vspace{8pt}

\end{document}